# Title: Anomalous normal state gap in an electron-doped cuprate

**Authors:** Ke-Jun Xu[1,2,3], Junfeng He[1,2,4], Su-Di Chen[1,2,3,5], Yu He[6], Sebastien N. Abadi[1,2,7], Costel. R. Rotundu[1,2], Young S. Lee[1,2,3], Dong-Hui Lu[8], Qinda Guo[9], Oscar Tjernberg[9], Thomas P. Devereaux[1,2,10], Dung-Hai Lee[5,11]*, Makoto Hashimoto[8]*, Zhi-Xun Shen[1,2,3,7]*

**Affiliations:**

[1] Stanford Institute for Materials and Energy Sciences, SLAC National Accelerator Laboratory, 2575 Sand Hill Road, Menlo Park, California 94025, USA

[2] Geballe Laboratory for Advanced Materials, Stanford University, Stanford, California 94305, USA

[3] Department of Applied Physics, Stanford University, Stanford, California 94305, USA

[4] Department of Physics and CAS Key Laboratory of Strongly-coupled Quantum Matter Physics, University of Science and Technology of China, Hefei, Anhui 230026, China

[5] Department of Physics, University of California, Berkeley, California 94720, USA

[6] Department of Applied Physics, Yale University, New Haven, Connecticut 06511, USA

[7] Department of Physics, Stanford University, Stanford, California 94305, USA

[8] Stanford Synchrotron Radiation Lightsource, SLAC National Accelerator Laboratory, 2575 Sand Hill Road, Menlo Park, California 94025, USA

[9] Department of Applied Physics, KTH Royal Institute of Technology, Hannes Alfvéns väg 12, 114 19 Stockholm, Sweden

[10] Department of Materials Science and Engineering, Stanford University, Stanford, California 94305, USA

[11] Material Sciences Division, Lawrence Berkeley National Laboratory, Berkeley, California 94720, USA

*Corresponding authors: D.-H. Lee (dunghai@berkeley.edu), M. H. (mhashi@slac.stanford.edu), Z.-X. S. (zxshen@stanford.edu)

**Abstract:**

In the underdoped n-type cuprate $Nd_{2-x}Ce_xCuO_4$, long-ranged antiferromagnetic order reconstructs the Fermi surface, resulting in a putative antiferromagnetic metal with small pockets. Using angle-resolved photoemission spectroscopy, we observe an anomalous energy gap, an order of magnitude smaller than the antiferromagnetic gap, in a wide range of the underdoped regime and smoothly connecting to the superconducting gap at optimal doping. After carefully considering all the known ordering tendencies in tandem with the phase diagram, we hypothesize that the normal state gap in the underdoped n-type cuprates originates from Cooper pairing. The high temperature scale of the normal state gap raises the prospect of engineering higher transition temperatures in the n-type cuprates comparable to that of the p-type cuprates.

**One sentence summary**: Angle-resolved photoemission spectroscopy reveals an anomalous gap in the underdoped n-type cuprates.

**Main Text:**

In the high-transition-temperature cuprate superconductors, the doping-temperature phase diagram has been intensely studied (*1*) to determine the various intertwined phases that are born from doping the parent Mott insulator. The electron-doped (n-type) side of the phase diagram was previously thought to be simpler than the hole-doped (p-type) side due to the dominance of antiferromagnetism (AF) (*2*), which reconstruct the Fermi surface (*3, 4, 5, 6*) to form an electron-like pocket centered at $(0, \pi)$ and a hole-like pocket centered at $(0.5\pi, 0.5\pi)$ (*6*). At low dopings, the large AF gap pushes the small hole pocket below $E_F$, resulting in an apparent Fermi surface consisting of only the $(0, \pi)$ electron pocket (*7*). It is thought that the pseudogap, which depletes the Brillouin zone boundary states in the p-type cuprates (*8, 9, 10, 11*), is largely absent in the n-type cuprates (*2*). Rather, the underdoped n-type cuprates are thought to be AF metals with reconstructed Fermi surfaces (*7*).

However, the validity of the AF metal ground state is cast into doubt by several earlier experimental findings. (i) Transport measurements have unveiled a prevalent upturn in resistivity at low temperatures (*12*) with a negative magnetoresistance persisting to high fields (*13*). Despite its prevalence, the resistivity upturn remains inadequately understood and has been ascribed to weak localization (*14*), Kondo effect (*13*), and scattering from AF puddles (*15*). However, it's important to emphasize that these proposed effects are not anticipated to induce the formation of a gap. (ii) In tunneling experiments conducted on superconducting n-type cuprate films, when the bulk superconductivity is suppressed using a magnetic field, a normal state gap of comparable magnitude to the superconducting gap becomes apparent (*16, 17, 18, 19*). However, the absence of momentum resolution and the limited availability of data in the non-superconducting underdoped regime give rise to conflicting interpretations regarding the nature of this observed gap (*18, 19*). (iii) Angle-resolved photoemission spectroscopy (ARPES) experiments, using the leading edge method, observed a low energy gap in extremely underdoped $Nd_{2-x}Ce_xCuO_4$ (NCCO) (*20*). Though, the broad spectrum only allowed the indirect observation of a small gap at dopings below x = 0.05, which is far from the doping level where bulk superconductivity occurs (x > 0.13 in NCCO).

Using a much-improved experimental set up and high-quality single crystals, we perform a comprehensive study of the low energy spectrum of NCCO across the entire accessible doping range. We discover a significant normal state gap (NSG) on the AF-reconstructed electron Fermi surface, that intriguingly evolves into the superconducting gap near optimal doping. Our characterization of momentum, doping, and temperature dependences excludes AF, charge order, and Coulomb gap as the origin of this NSG. The exhaustion of alternatives compels us to propose that NSG represents a pairing gap with a temperature scale up to ~140 K. Furthermore, this hypothesis offers a comprehensive elucidation that establishes a coherent link between the presence of a gapped state and the aforementioned anomalous transport properties. In the ensuing discussions, we present arguments explaining why the system would behave as a weak insulator with a negative magnetoresistance rather than a bulk superconductor despite the presence of pairing, and why the temperature at which the gap fills approximately coincides with the temperature at which a resistivity minimum is observed in transport measurements.

## Electronic structure of $Nd_{1.89}Ce_{0.11}CuO_4$

We commence by examining the overall electronic structure of NCCO at x = 0.11, wherein long-range AF order is present (*21*). The sample is not bulk superconducting, as shown in the magnetic susceptibility (supplementary Fig. S1). Figure 1a illustrates the Fermi energy mapping at 8 K. Notably, states within a 20 meV window of the Fermi energy ($E_F$) are predominantly governed by the reconstructed electron pocket centered at $(0, \pi)$. Further insights into the energy-momentum cuts 1-3 are presented in Figure 1B-D, with the corresponding energy distribution curves (EDCs) depicted in Fig. 1F-H. Near the Brillouin zone diagonal, the maximum of the reconstructed valence band lies approximately 50 meV below $E_F$ (Figure 1F). This energy separation is the part of the AF gap ($\Delta_{AF}$) below the Fermi energy. Due to the absence of a well-defined dispersion within $\Delta_{AF}$, it is not possible to define a Fermi momentum ($k_F$) when $\Delta_{AF}$ straddles $E_F$. For ease of reference, we employ the peak of the momentum distribution curve (MDC) at $E_F$ ($k_{peak}$), arising from a small amount of intensity leaking into $\Delta_{AF}$, to define the position of the underlying unconstructed large Fermi surface. As $k_{peak}$ approaches the reconstructed pocket, the center of $\Delta_{AF}$ shifts from above the Fermi energy to the Fermi energy and ultimately below it. Near the Brillouin zone boundary (Fig. 1D and 1H), the dispersion demonstrates an apparent Fermi level crossing, allowing the identification of $k_F$. The spectra along $k_{peak}$ and $k_F$ are summarized in Fig. 1E.

## Momentum dependence

With an understanding of the overall electronic structure, we examine the low energy spectra. Fig. 2A shows the symmetrized EDCs at $k_{peak}$ and $k_F$ as a function of the Fermi surface angle θ (defined in inset of Fig. 2A) at 8 K. Within this graph, we can discern two distinct features: one characterized by binding energy ($E_B$) ranging from approximately 50 to 100 meV (depicted by grey dots), which corresponds to the portion of $\Delta_{AF}$ situated below the Fermi energy, and another at an $E_B$ of approximately 10 meV (represented by red dots). This is the gap we denoted as $\Delta_{NSG}$. We note that the spectra are symmetrized to visualize the low energy gap. $\Delta_{AF}$ is not particle-hole symmetric except at the AF hot spots, whereas $\Delta_{NSG}$ appears to be particle-hole symmetric (see inset of Fig. 2C, as well as supplementary Figs. S2-4). The energies associated with these two gaps are summarized in Fig. 2B as a function of θ. In this case, the scale for the dispersive $\Delta_{AF}$ (grey data points) is represented on the left axis, while the scale for $\Delta_{NSG}$ (red data points) with little angular dependence is displayed on the right axis. Here we note the approximately one order of magnitude difference between the two gap energy scales and the different angular dependences.

## Temperature dependence

With increasing temperature, $\Delta_{NSG}$ fills and becomes undetectable at around 50 K for x = 0.11 (Fig. 2C). Here, due to possible fluctuations of the gap order parameter that may extend to higher temperatures but not easily observable in the spectra, we define the gap filling temperature scale $T_{fill}$ as the temperature at which no distinct double peaks can be discerned in the symmetrized spectra. At $x = 0.11$, $T_{fill}$ is approximately 50 K. Such a gap-filling temperature scale approximately coincides with the upturn temperature scale in resistivity measurements of an $x = 0.1$ single crystal sample from Ref. (24) (Fig. 2D). The consistent scaling of the inverse low energy gap spectral weight depletion and the resistivity upturn below $T_{fill}$ is also shown in Fig. 2D, suggesting that the $E_F$ density-of-state depletion is the cause of the resistivity upturn. In addition to the symmetrized temperature-dependent EDCs, we also show the Fermi-Dirac-corrected spectra and EDCs near the Brillouin zone boundary in supplementary Figs. S3-4.

**Doping dependence**

We further study the doping dependence of the low energy spectrum to understand the evolution of NSG. Fig. 3A and 3B show the symmetrized low temperature spectra and $k_F$ EDCs near the Brillouin zone boundary from $x = 0.04$ to $x = 0.19$. The gap at the Brillouin zone boundary decreases monotonically from 38 meV at $x = 0.04$ to approximately 0 at $x = 0.19$. The most overdoped sample ($x = 0.19$, estimated from the Fermi surface volume) is accessed through surface K dosing, as bulk crystals of NCCO cannot be chemically doped beyond $x = 0.17$ due to the Ce solubility limit. As we observe a small but well-defined coherence peak in the $x = 0.15$ (22) and $x = 0.17$ spectra, we conclude that the Brillouin zone boundary gap at this doping is of superconducting origin. In the underdoped regime where the system is not a bulk superconductor, we phenomenologically label the gap as the NSG. Remarkably, $\Delta_{NSG}$ evolves smoothly to the $\Delta_{SC}$ at $x = 0.15$. Importantly, the shape and sharpness of the gap for the $x = 0.11$ and $x = 0.08$ samples show strong resemblance to a superconducting gap. The sharp gap edges at low dopings also render it unlikely that the NSG arises from Coulomb localization (see also supplementary text). Even at the low doping of $x = 0.04$, the zone boundary spectra host a reasonably well-defined gap edge at 15 K, and the gap fills up at around 140 K (supplementary Fig. S5). However, we note that we cannot rule out Coulomb localization as a component of the gap, especially at low dopings. Additional spectra of the underdoped samples are provided in supplementary Figs. S6-7.

**Phase diagram and phenomenology**

The discovery of the NSG in the underdoped n-type cuprates motivates us to add this anomalous regime to the phase diagram. The doping evolution of the Brillouin zone boundary gap and $T_{fill}$ is shown in Fig. 4A. The regime where the NSG exists is marked by the red shaded area. One immediate question is how the NSG evolves to the Mott insulator at half filling? This is resolved by considering the results from a previous study using precision surface Rb dosing of the Mott insulator $Ca_3Cu_2O_4Cl_2$ (23), where a notable chemical potential jump occurs at a small electron doping of less than 1%, concomitant with the emergence of a Fermi surface around $(0, \pi)$ within the Mott gap. Importantly, the NSG discovered in this paper occurs on this $(0, \pi)$ Fermi surface, and hence does not smoothly evolve into the Mott gap at half filling.

This phase diagram also displays $T_{min}$, the temperature at which the resistivity reaches a minimum, as yellow data points (13, 24, 25). Furthermore, the bulk superconducting $T_c$ (24), and temperature scales associated with AF (*21, 26, 27*) and charge density wave (CDW) (*28, 29*) are marked by the blue, gray, and green symbols, respectively. From this plot we observe the following: (i) $T_{fill}$ roughly follows the resistivity minimum temperature in the underdoped samples (Fig. 4A); (ii) The onset temperature and energy scale of the AF pseudogap are an order of magnitude larger than $T_{fill}$ and $\Delta_{NSG}$; (iii) The onset temperature of the CDW correlations (as measured by resonant X-ray scattering, or RXS) is much higher than $T_{fill}$ at most dopings, and its doping trend is opposite to that of $T_{fill}$.

To gain insights into the origin of the NSG, we examine its phenomenology in the context of known ordering tendencies. We first consider the possibility of AF causing the NSG. The compelling evidence of significant differences in energy and onset temperature scales (Fig. 4A) between $\Delta_{NSG}$ and AF correlation, along with the observation that $\Delta_{NSG}$ forms on the AF-reconstructed pockets (Fig. 2A), allows us to confidently conclude that AF is not the direct cause of $\Delta_{NSG}$.

As the reconstructed electron pocket hosts parallel straight sections at x = 0.11 that is conducive to Fermi surface nesting (Supplementary Fig. S8), we consider the possibility of CDW inducing the NSG. Comparing the behavior of the NSG to the CDW signal from RXS (*28, 29*), we find a completely opposite doping trend and distinct temperature scale between $\Delta_{NSG}$ and the CDW. This strongly suggests that the CDW does not directly cause the NSG. In addition, it's important to consider the changes in the shape of the electron pocket at low dopings (Supplementary Fig. S9). The observation that the electron pocket becomes more rounded at low dopings indicates a weakening of the nesting condition, while the NSG grows stronger at low dopings. This finding further strengthens the argument against a nesting-induced CDW being the direct cause of $\Delta_{NSG}$.

**Microscopic origins of the normal state gap**

Given the smooth evolution of $\Delta_{NSG}$ to $\Delta_{SC}$ and the particle-hole symmetric nature of $\Delta_{NSG}$, we conjecture that the NSG is a pairing gap. Regarding the pairing symmetry, we first transform the real space pairing order parameter into momentum space. The simplest examples are *s*-wave $\Delta(k) = \Delta_0$, extended *s*-wave $\Delta(k) = \Delta_0\big(\cos k_x + \cos k_y\big)$, and *d*-wave $\Delta(k) = \Delta_0\big(\cos k_x - \cos k_y\big)$. We then solve the Bogoliubov-de Gennes equation with both the antiferromagnetic and superconducting order parameters. Our findings revealed distinct behaviors for each symmetry: the d-wave pairing exhibits an approximately isotropic gap around the Fermi surface, consistent with the experimental findings in Fig. 2B, while the s-wave pairing symmetries produced gap nodes (supplementary Fig. S10). We have checked the validity of the gap symmetries as we extend the pairing to more distant neighbors.

The next question we address is the pairing mechanism. It is evident that spin waves cannot mediate pairing in the long-ranged AF ordered state (*30*). However, short-range AF super-exchange interaction can account for the d-wave pairing (*31*). The observed isotropic gap around the reconstructed Fermi surface is consistent with d-wave pairing mediated by super-exchange. On the other hand, the apparent competition between the NSG and CDW in the doping-temperature phase diagram (Fig. 4A) suggests that the NSG may arise from a pairing mechanism that drives both CDW and pairing (Fig. 4B-C). One possibility is electron-phonon interaction, which has been studied in the context of CDW and pairing in the cuprates (*32, 33, 34*). It is also known that electron-phonon interactions can be amplified by short-range antiferromagnetic correlations (*35, 36*). The importance of electron-phonon interactions is also highlighted by the recent discovery of a next-nearest-neighbor phonon-induced attractive interaction in the chain cuprates (*37*). Thus, the phonons that scatter the electrons between the parallel segments of the Fermi surface can drive both the CDW and d-wave pairing (Fig.4B-C). We note that for such electron-phonon scattering, the d-wave and s-wave pairing channels are degenerate. However, in the presence of Hubbard-like repulsive electron-electron interaction, the d-wave channel would be energetically favorable.

**Absence of global coherent superconductivity**

Another important question is why the resistivity increases as the temperature drops below $T_{fill}$. To address this question, we consider the effects of disorder in a low-carrier-density superconductor. It was previously shown that non-magnetic scattering of the Bogoliubov quasiparticles from one gap sign to the opposite gap sign can break up Cooper pairs and result in granular superconductivity (*38*). According to this physical picture (supplementary Fig. S11), the direct tunneling of Cooper pairs between the superconducting islands is significantly weakened when the superconducting island becomes sparse. As a result, the main contribution to conduction in a disordered *d*-wave superconductor arises from the quantum diffusion of the unpaired carriers within the metallic background. When the temperature falls below $T_{fill}$, a depletion of the population of unpaired carriers occurs, resulting in an increase in resistivity as observed experimentally.

The granular superconductivity picture explains other aspects observed in the ARPES data: The lack of a superconducting coherence peak at the edge of $\Delta_{NSG}$ (Fig. 3A) and the filling up of a gap with a fixed magnitude (Fig. 2C), are expected for a phase incoherent superconductor. This picture is also consistent with the negative magnetoresistance and suppression of the resistivity upturn observed at high fields (*13, 39*) – the high magnetic fields can break Cooper pairs, resulting in an increased population of unpaired carriers for metallic diffusion. The NSG observed in underdoped NCCO is reminiscent of the putative incoherent pairing gap in disordered $InO_x$ films (*40*), which is thought to host localized Cooper pairs on the insulating side of the superconductor-insulator transitions and shows similar negative magnetoresistance at high fields (*41*).

**Outlook**

Our results here reveal that the n-type cuprates host a distinct normal state gap spanning a wide range in the underdoped regime. If this gap is indeed a pairing gap, it holds exciting implications. The high temperature scales and large gap energies observed for $\Delta_{NSG}$ suggest the potential for high-temperature superconductivity comparable to the maximum transition temperatures observed in the p-type cuprates (*42*). This raises the possibility that the lack of higher $T_c$ superconductivity in n-type cuprates is primarily attributed to the limited mobility of Cooper pairs. Further experimental and theoretical works are expected to elucidate the order parameter symmetry and microscopic mechanism of pairing on the reconstructed Fermi surface.

**Acknowledgments:**

We thank S. A. Kivelson for fruitful discussions. Use of the Stanford Synchrotron Radiation Lightsource, SLAC National Accelerator Laboratory, is supported by the U.S. Department of Energy, Office of Science, Office of Basic Energy Sciences under Contract No. DE-AC02-76SF00515. **Funding:** Use of the Stanford Synchrotron Radiation Lightsource, SLAC National Accelerator Laboratory, is supported by the U.S. Department of Energy, Office of Science, Office of Basic Energy Sciences under Contract No. DE-AC02-76SF00515. D.H.Lee was funded by the US Department of Energy, Office of Science, Office of Basic Energy Sciences, Materials Sciences, and Engineering Division under Contract No. DEAC02- 05-CH11231. **Author contributions:** K.-J.X. and Z.-X.S. conceived the experiment. K.-J.X., J.H., S.A., C.R.R., and Y.S.L synthesized the samples. K.-J.X. M.H., and D.-H.Lu performed the ARPES measurements at SSRL. K.-J.X., J.H., S.-D.C., Y.H., Q.G., O.T., D.-H.Lee, M.H. and Z.-X.S. analyzed and interpreted the ARPES data. D.-H.Lee conceived the theoretical model. K.-J.X., M.H., D.-H.Lee and Z.-X.S. wrote the manuscript with input from all authors. **Competing interests:** The authors declare no competing interest. **Data and materials availability:** All data presented in this paper are available at Stanford Digital Repository.

**Supplementary Materials**

Materials and Methods

Supplementary Text

Figs. S1 to S14

References 43-48

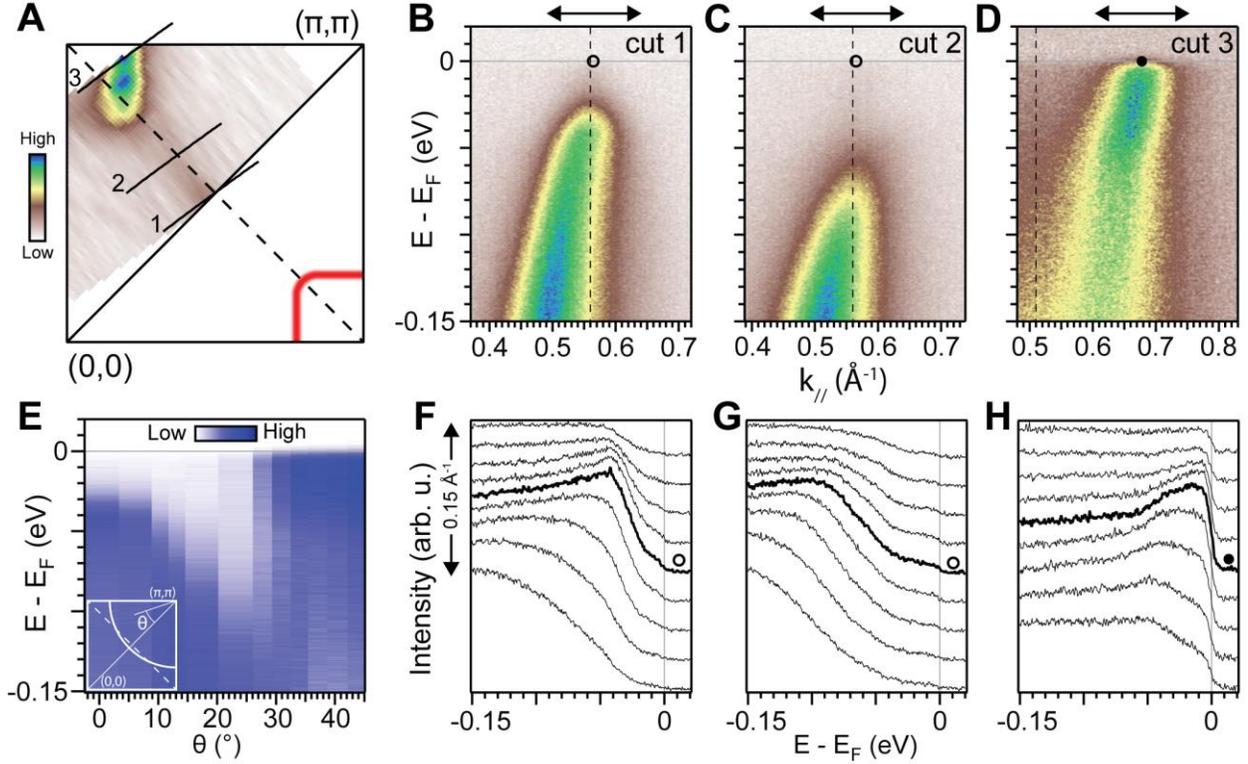

**Figure 1. Electronic structure of Nd$_{1.89}$Ce$_{0.11}$CuO$_4$.** (**A**) Top left: constant energy mapping constructed using intensity from ±10 meV of the Fermi energy ($E_F$). Bottom right: schematic showing the reconstructed electron pocket. (**B-D**) Energy-momentum cuts corresponding to the numbered cuts in (A). (**E**) color plot of the spectra at $k_{peak}$ (see below) as a function of the Fermi surface angle θ (see inset). The spectra are normalized by the angular-dependent matrix element extracted from an overdoped sample with a simple large circular Fermi surface. (**F-H**) Energy distribution curves (EDCs) of the cuts in (B-D) in a range of 0.15 Å$^{-1}$ centered at the momentum indicated by the open or solid circles. Despite a nearly full gap at the Brillouin zone diagonal and hot spot, there is a small number of states (<5% of the AF peak feature height) leaked into the AF gap that we use to define a $k_{peak}$, which roughly tracks the AF zone boundary instead of the underlying $k_F$. Here, $k_{peak}$ is indicated by open circles for cuts 1-2 and the Fermi momentum $k_F$ on the reconstructed pocket is indicated by the solid circle for cut 3. Grey vertical line indicates $E_F$. Measurement temperature for all panels is 8 K.

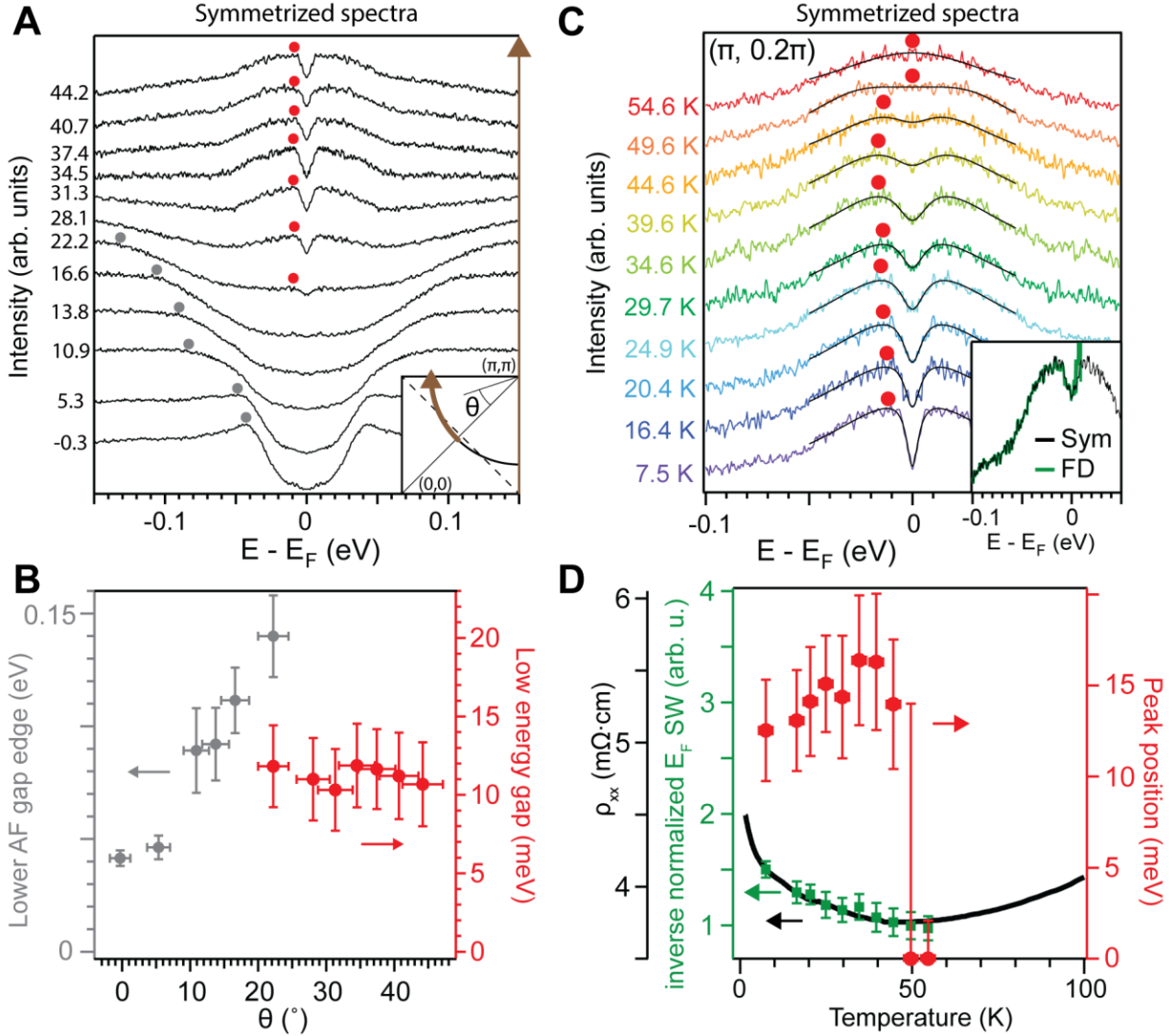

**Figure 2. Low energy spectrum in Nd$_{1.89}$Ce$_{0.11}$CuO$_4$.** (**A**) Spectra at k$_F$ and k$_{peak}$ as a function of the Fermi surface angle θ (see inset). Grey dots indicate the AF feature, red dots indicate the normal state low energy gap (Δ$_{NSG}$) on the reconstructed electron pocket. The spectra are normalized by the angular-dependent matrix element extracted from an overdoped sample with a simple large circular Fermi surface. (**B**) Angular dependence of the AF feature (grey) and Δ$_{NSG}$ (red). Note the different left and right axes. Measurement temperature in (A) and (B) is 8 K. The spectra in A and B are symmetrized for visualizing the gap. We note that the AF gap is not particle-hole symmetric. Details on the particle-hole symmetry nature of the low energy gap are presented in Supplementary Figs. S2-4. (**C**) Temperature dependence of Δ$_{NSG}$ at the Brillouin zone boundary k$_F$. Black line on top of the data points are low energy fits using a phenomenological model (see methods). Inset: symmetrized (black) and Fermi-Dirac-corrected (green) spectra for 34.6 K, showing that the gap appears to be particle-hole symmetric. (**D**) Temperature dependence of the inverse spectral weight at E$_F$ (left axis, green data points), low energy gap peak position (right axis, red data points), and resistivity from an x = 0.1 NCCO single crystal (*24*) (left axis, black curve).

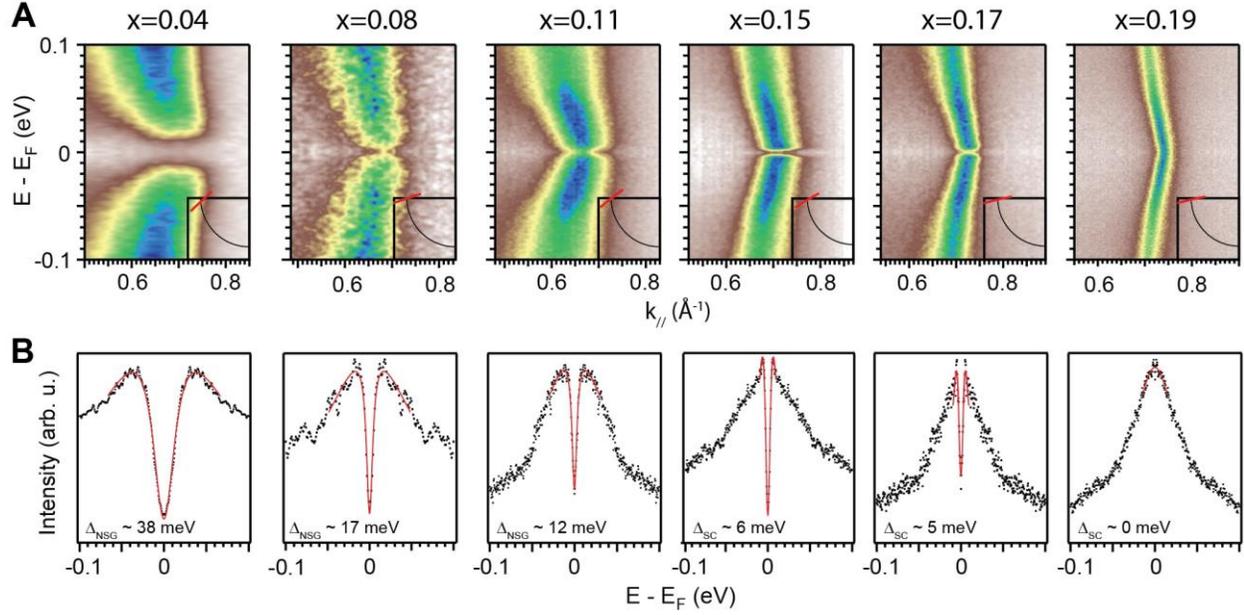

**Figure 3. Doping dependence of the low energy spectrum.** (**A**) Symmetrized energy-momentum spectra at the Brillouin zone boundary $k_F$ (see insets for cut location) for different dopings. Spectra are symmetrized with respect to $E_F$ to visualize the gap. The insets show schematics of the underlying large Fermi surface for uniformity, but we note that the underdoped regime has reconstructed small electron pockets. The x = 0.19 sample is obtained by surface K dosing (see methods), with the doping level estimated by the Fermi surface volume. The slightly different cut directions in A are due to the mounting orientations of the different samples. (**B**) Doping dependence of the symmetrized EDC at $k_F$ (black data points) with a phenomenological fit of the low energy data range (red curve). Measurement temperatures in (A) and (B) are 8 K, except for the x = 0.04 data, which is taken at 15 K. Here, the EDCs are symmetrized to visualize the gap. Additional details on the particle-hole symmetry nature of the low energy gap are presented in Supplementary Figs. S2-4.

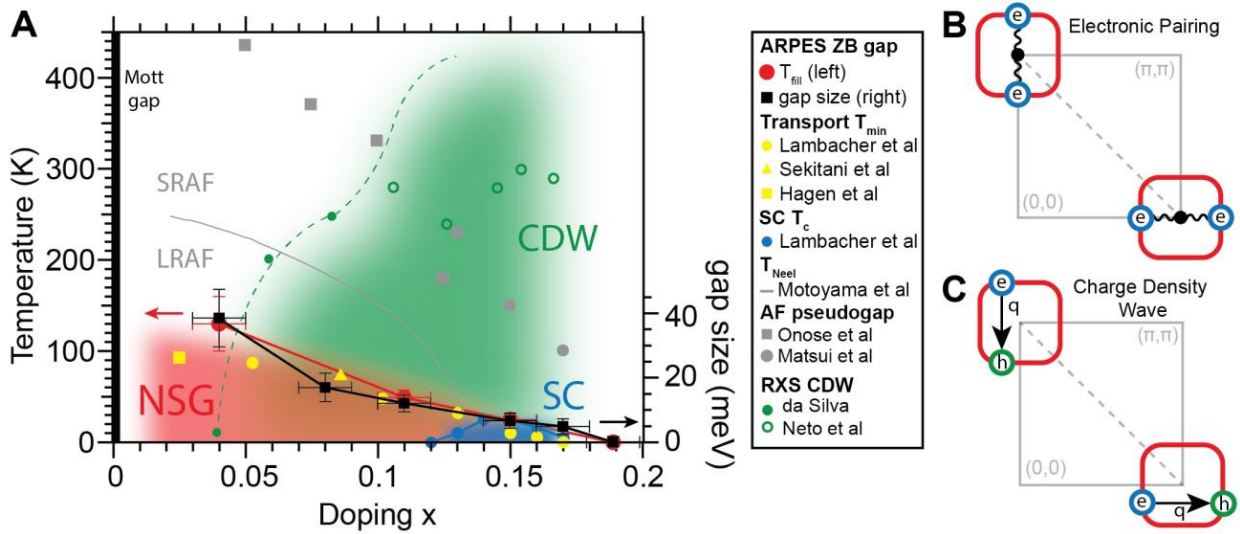

**Figure 4. Phase diagram of Nd$_{2-x}$Ce$_x$CuO$_4$.** (**A**) Doping dependence of the zone boundary gap size (black squares, right axis) and T$_{fill}$ (red circles, left axis), along with the temperature scales of other phases in Nd$_{2-x}$Ce$_x$CuO$_4$ (see legend). Yellow data points (left axis) show the resistivity upturn temperature scale, defined by the minimum resistivity temperature, in NCCO crystals and films (*13, 24, 25*). The resistivity upturn data points within the superconducting dome are obtained by suppressing bulk superconductivity with a magnetic field. Blue data points (left axis) indicate the superconducting transition temperatures obtained from magnetic susceptibility (*24*). Grey solid line (left axis) is the Néel temperature from neutron scattering (*21*). Grey data points (left axis) show the AF pseudogap temperatures from optical spectroscopy (grey squares) (*26*) and ARPES (grey circles) (*27*). Green data points (left axis) show the charge density wave (CDW) temperature from resonant X-ray scattering (RXS) (*28*): filled circles indicate the signal disappearance temperature; open circles indicate the signal saturation temperature. Green dashed line highlights the approximate boundary of the CDW, adapted from (*29*). SRAF refers to short-ranged antiferromagnetism, LRAF refers to long-ranged antiferromagnetism, CDW (green region) refers to charge density wave, NSG (red region) refers to the normal state low energy gap, and SC (blue region) refers to bulk superconductivity. The thick black vertical line near 0 doping indicates the presence of the full Mott gap (see discussion in text). (**B**) Schematic showing pair formation on the reconstructed electron pocket. (**C**) Schematic showing Fermi surface nesting leading to CDW on the parallel straight sections of the reconstructed electron pocket.

# Supplementary Materials for

## Anomalous normal state gap in an electron-doped cuprate

Ke-Jun Xu, Junfeng He, Su-Di Chen, Yu He, Sebastien N. Abadi, Costel. R. Rotundu, Young S. Lee, Dong-Hui Lu, Qinda Guo, Oscar Tjernberg, Thomas P. Devereaux, Dung-Hai Lee, Makoto Hashimoto, Zhi-Xun Shen

Corresponding author: D.-H. Lee (dunghai@berkeley.edu), M. H. (mhashi@slac.stanford.edu), Z-X. S (zxshen@stanford.edu)

**The PDF file includes:**

Materials and Methods
Supplementary Text
Figs. S1 to S14
References 43-48

**Materials and Methods**

*Sample synthesis and annealing*

$Nd_{2-x}Ce_xCuO_4$ (NCCO) single crystals were grown with the traveling-solvent floating-zone method with CuO flux in the molten zone. The crystals were annealed at 900° C under flowing Ar gas. The magnetic susceptibility properties of the samples were characterized in a Physical Property Measurement System using the AC susceptibility method. For the superconducting samples (x = 0.15 and x = 0.17), the onset of diamagnetism is taken as $T_c$, as that is usually the temperature where the resistivity reaches 0. The Ce content of the crystals was characterized by wavelength dispersive spectroscopy in an electron probe microanalyzer.

*ARPES measurements*

ARPES measurements were performed at beamlines 5-2 and 5-4 (Stanford Synchrotron Radiation Lightsource, or SSRL). Single crystal NCCO samples were mounted on top of copper posts with H20E silver epoxy and Torrseal. Laue back scattering was performed to align the sample in the basal plane. A ceramic top post was mounted with Torrseal for in situ cleaving. 53 eV photons were used for the measurements at 5-2 and 16.5 eV photons were used for the measurements at 5-4. The beam spot size is estimated to be around 30 μm by 10 μm at 5-2 and about 100 μm by 150 μm at 5-4. The base pressure is about $4.0 \times 10^{-11}$ Torr at 5-2 and $1.8 \times 10^{-11}$ Torr at 5-4. For the temperature dependent measurements, we vary the sample temperature with a local heater on the sample stage to avoid sample aging. This way, the rest of the manipulator arm is kept at a low temperature to reduce outgassing of adsorbents.

The Fermi level $E_F$ is measured on a reference polycrystal gold and carefully checked for extrinsic sample charging and space charging effects for the low energy gap measurements (supplementary Fig. S12). At 5-4, $E_F$ is generally measured about every 30-45 minutes to account for a slow drift of the photon energy on the scale of < 0.5 meV/hour due to monochromator warm up and diurnal variations. We note that once the beamline reaches a stable state, the $E_F$ drift at 5-4 is usually about 0.1 meV/hour or less. The experimental resolution for the measurements at 5-2 is about 8 meV and that at 5-4 is about 4 meV, both of which are determined from fitting the reference gold Fermi cutoff. The temperature dependence of the gap is checked by thermal cycling (supplementary Fig. S13), ensuring that the gap filling is intrinsic.

*Surface potassium dosing*

Bulk NCCO samples cannot be chemically doped to beyond x = 0.17. To access the extremely overdoped regime where there are no visible signatures of AF, surface K dosing was performed to introduce additional electrons into the sample surface layer. Due to the surface sensitive nature of the ARPES measurements, introducing surface dopants is a suitable method to probe the intrinsic doping dependence of the sample. The surface K dosing was performed using a SAES alkali metal dispenser located in the measurement chamber but not in the line of sight of the electron analyzer. The evaporation was performed at 5.5A with 30s doses at a time. The pressure of the chamber reaches a maximum of about $1.5 \times 10^{-10}$ Torr during the evaporation. The reference polycrystal gold was covered during the evaporation to ensure the gold surface remains clean and no extrinsic shift of the reference Fermi level is induced by dopants.

While high levels of surface dopants can degrade the ARPES spectra, the fact that the spectra become sharper after surface K dosing (Fig. 3A) indicates that no significant spectra degradation occurs at the dosing levels used in our experiments.

### ARPES data processing

To properly process the raw data to the presented data in the figures, several careful calibrations and conversions are required. Standard ARPES data processing procedures were performed (*43, 44*), including: detector channel inhomogeneity calibration, detector nonlinearity calibration (supplementary Fig. S14), analyzer slit inhomogeneity and lensing calibration, $E_F$ calibration with respect to an electrically connected polycrystalline gold reference, and emission-angle-to-momentum conversion. Normalizations of ARPES spectral intensities are performed using the high-order light background intensity above $E_F$, well above any significant thermal tail of the Fermi-Dirac distribution. To remove the incoherent scattering background, a reference EDC background far away from dispersive features is subtracted from the spectra.

### Phenomenological fitting of the low energy gap

We perform fits of the low energy gap using the phenomenological model in (*45*), where the self-energy has the following form

$$\Sigma(k, \omega) = -i\Gamma_1 + \frac{\Delta^2}{\omega + \epsilon(k) + i\Gamma_0}$$

Here, $\Gamma_1$ is the single particle scattering rate, $\Delta$ is the gap magnitude, and $\Gamma_0$ is the inverse quasiparticle lifetime.

### Pairing gap calculation

Pairing is implemented in real space with nearest neighbor pairing. The interaction is given by

$$H(k_x, k_y) = \begin{pmatrix} E_k & \Delta_k & m & 0 \\ \Delta_k & -E_k & 0 & m \\ m & 0 & E_{k+q} & \Delta_{k+q} \\ 0 & m & \Delta_{k+q} & -E_{k+q} \end{pmatrix}$$

Where $E_k$ is the tight binding band dispersion given by

$$E_k = \mu - 2t'(\cos k_x + \cos k_y) - 4t''(\cos k_x \cos k_y)$$

We have used $\mu = -0.038847$, $t' = 0.23428$, $t'' = -0.093823$.

$m$ is the AF order parameter that shapes the reconstructed electron pockets. We have used $m = 0.22$ in this calculation. $\Delta_k$ is the gap function, which is given below for the respective symmetries:

| Simple s-wave | $\Delta_0$ |
|---|---|
| Extended s-wave | $\Delta_0(\cos k_x + \cos k_y)$ |

| d-wave | $\Delta_0 (\cos k_x - \cos k_y)$ |
|---|---|

We have used $\Delta_0 = 0.05$ for the extended s-wave calculation and $\Delta_0 = 0.02$ for the d-wave and simple s-wave calculations.

## Supplementary Text

### *Arguments disfavoring Coulomb gap as the origin for the normal state gap*

One possible scenario for the NSG is the disorder-induced Coulomb gap. However, it is unnatural for a Coulomb gap to smoothly evolve into the superconducting gap with increasing doping. The relatively well-defined spectral peaks at the NSG gap edge, even at low dopings, are also not consistent with a disordered system. Furthermore, because the NSG and resistivity upturn likely arise from the same physics, several properties of the transport disfavor the Coulomb gap scenario: (i) the resistivity does not exhibit any localization behavior for temperatures above $T_{fill}$, which is not typical for variable range hopping system. (ii) At high magnetic fields of order 50 T, the resistivity upturn is suppressed and negative magnetoresistance is observed. Such a behavior is not consistent with a Coulomb gap, as wave function confinement effects produce a positive magnetoresistance at high fields (*46*). We note that while negative magnetoresistance at low magnetic fields is associated with weak localization, such purely disorder-driven localization processes do not open a single particle spectral gap (*47*).

### *Inadequacy of previous explanations for the resistivity upturn*

Previous works on the resistivity upturn in the electron doped cuprates have suggested several candidate explanations, including weak localization, Kondo effect, and scattering from AF droplets. We provide explanations below for why these scenarios are not compatible with our ARPES observations.

1. Weak localization. Weak localization arises from quantum interference around disorder or impurities. Disorder-driven localization processes, in the absence of electron-electron interactions, drives localization through opening a mobility gap with a pure point spectrum. The single particle spectrum, which is measured by ARPES, would not exhibit a gap. Thus, weak localization is incompatible with our observations.

2. Kondo effect. Scattering of conduction electrons by magnetic impurities can give rise to a positive correction to the resistivity at low temperatures (*48*). However, for a spectral gap to open, there needs to be a substantial number of impurities as in the case of Kondo lattice systems. In such a case, the impurity band should be visible, but this is not the case in the ARPES spectrum. For low impurity concentrations, there is no coherent hybridization and there would be no single particle gap.

3. Scattering from AF "droplets". In this scenario studied in reference 15, disorder can induce local magnetic regions in a correlated metal. The additional magnetic scattering can lead to a positive resistivity correction. We note that this theory is suitable for samples with short-ranged AF order such as optimally doped samples and slightly underdoped samples, and cannot account for the gap observed in underdoped samples where long-ranged AF order is observed. Furthermore, scattering by dilute local magnetic moments is not expected to open a spectroscopy gap.

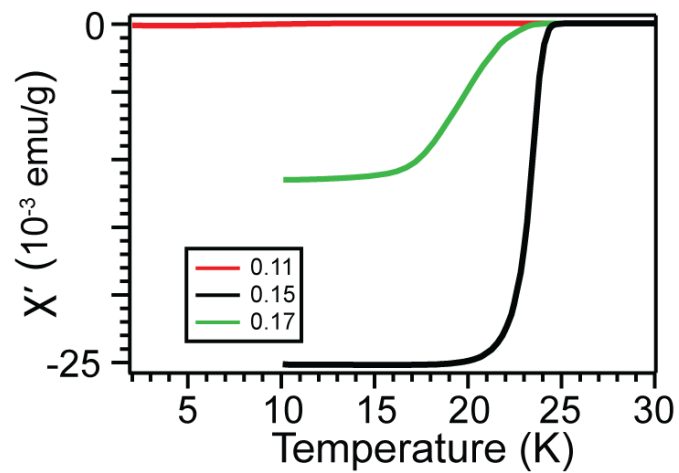

**Fig. S1.**

**Magnetic susceptibility curves of Nd$_{2-x}$Ce$_x$CuO$_4$.** Different colored curves represent the real part of the AC susceptibility for different dopings as indicated by the legend. Curves are measured with zero-field cooling. The excitation AC field is 5 Oe at a frequency of 4000 Hz. Curves are normalized by the normal state susceptibility above 25 K.

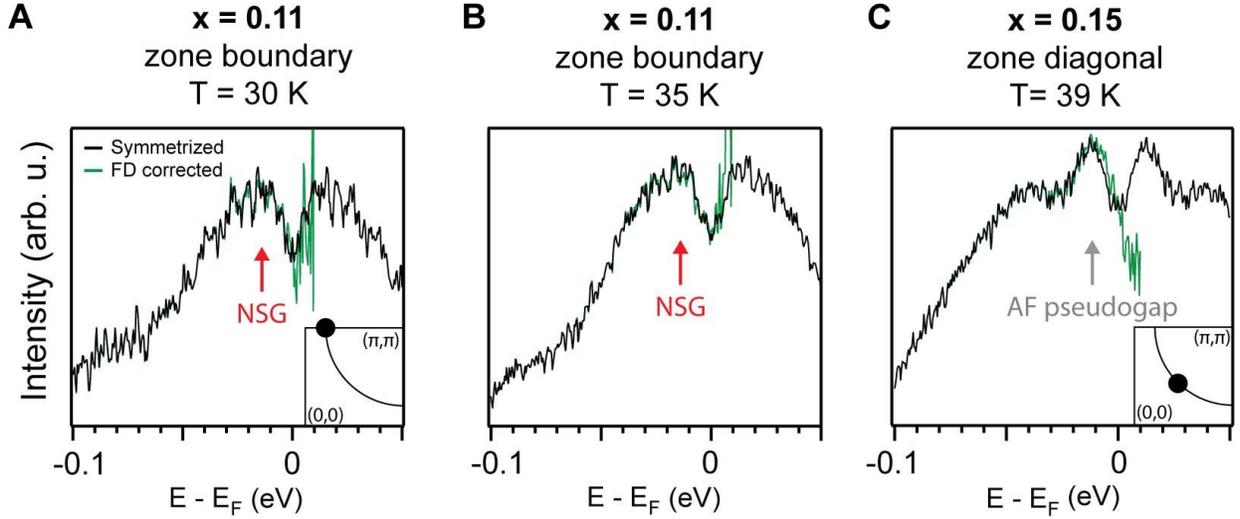

**Fig. S2.**

**Particle-hole symmetry of the low energy gap. (A-B)** EDCs at the Brillouin zone boundary (see inset) of an x = 0.11 sample, taken at 30 K (A) and 35 K (B). The black curve is the symmetrized data, and the green curve is the Fermi-Dirac-corrected data. The approximate agreement in the region above $E_F$ indicates that the gap here is particle-hole symmetric. **(C)** Reference spectra at the zone diagonal of an x = 0.15 sample, taken at 39 K. Here, the gap originates from short-range antiferromagnetism and is not particle-hole symmetric. As the measurement temperature is significantly above the bulk $T_c$ of 25 K, this gap is purely from the antiferromagnetic pseudogap. The difference between the symmetrized data and the Fermi-Dirac-corrected data in the region above $E_F$ confirms the particle-hole asymmetry. We note that the low energy gap is incipient, i.e. the spectral suppression is not complete. Therefore, despite the observed particle-hole gap symmetry, there are still residual electrons that can give rise to a finite hall signal.

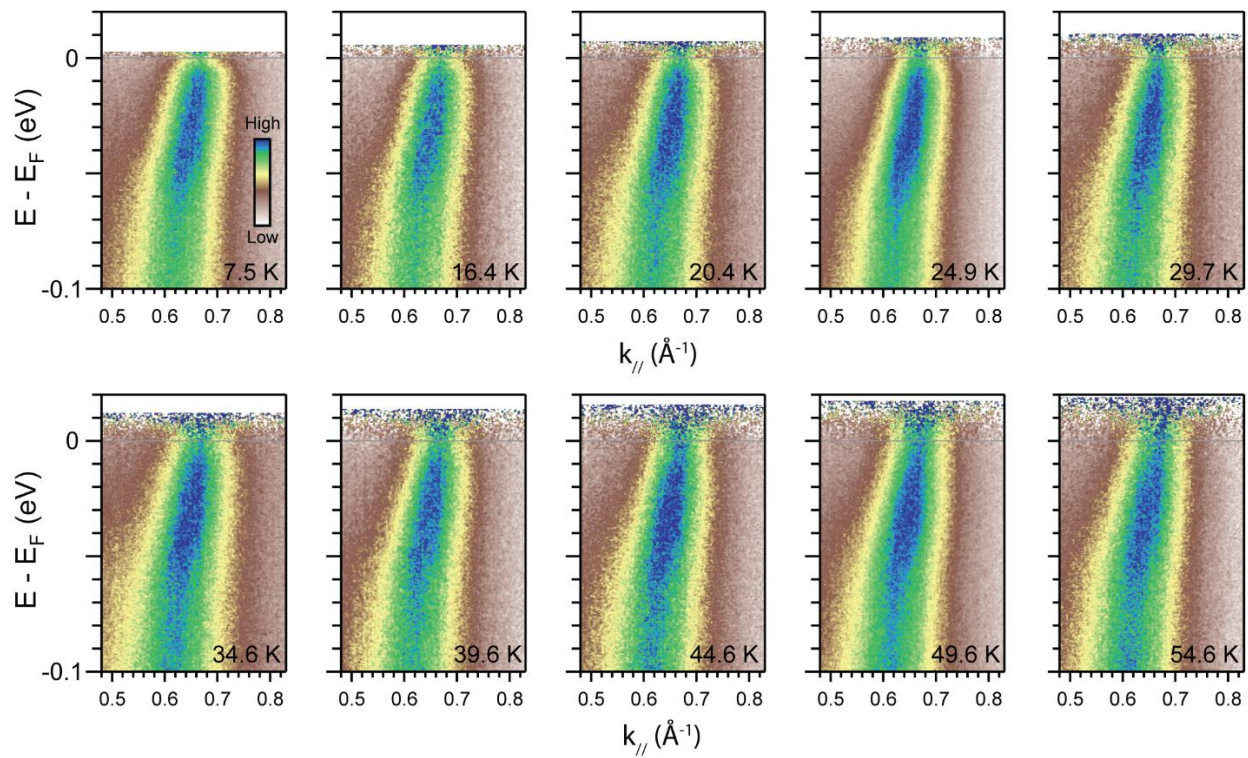

**Fig. S3.**
**Raw Fermi-Dirac-corrected spectra for the x = 0.11 sample.** The temperature of each spectrum is labeled at the bottom right of each panel. The grey horizonal line indicates the Fermi energy ($E_F$).

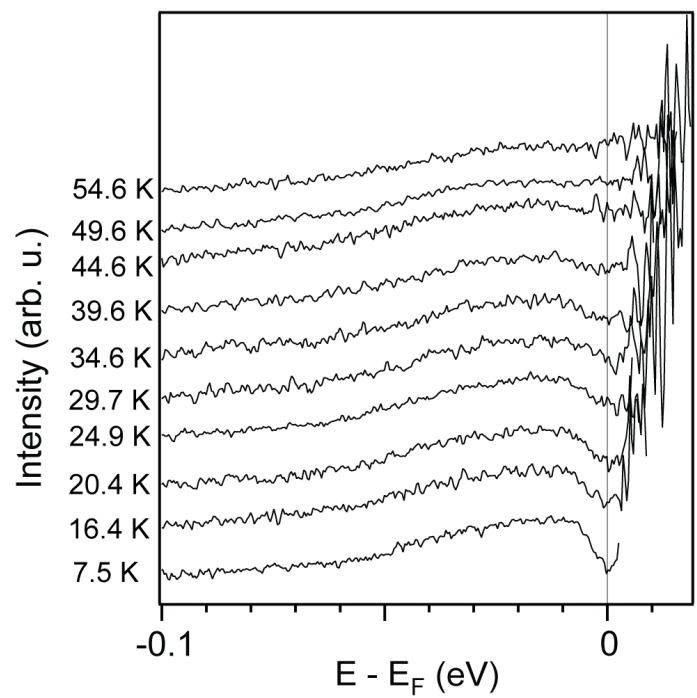

**Fig. S4.**
**Raw Fermi-Dirac-corrected energy distribution curves at $k_F$ for the x =0.11 sample.** The temperature of each EDC is labeled to the left. Grey vertical line indicates the Fermi energy ($E_F$).

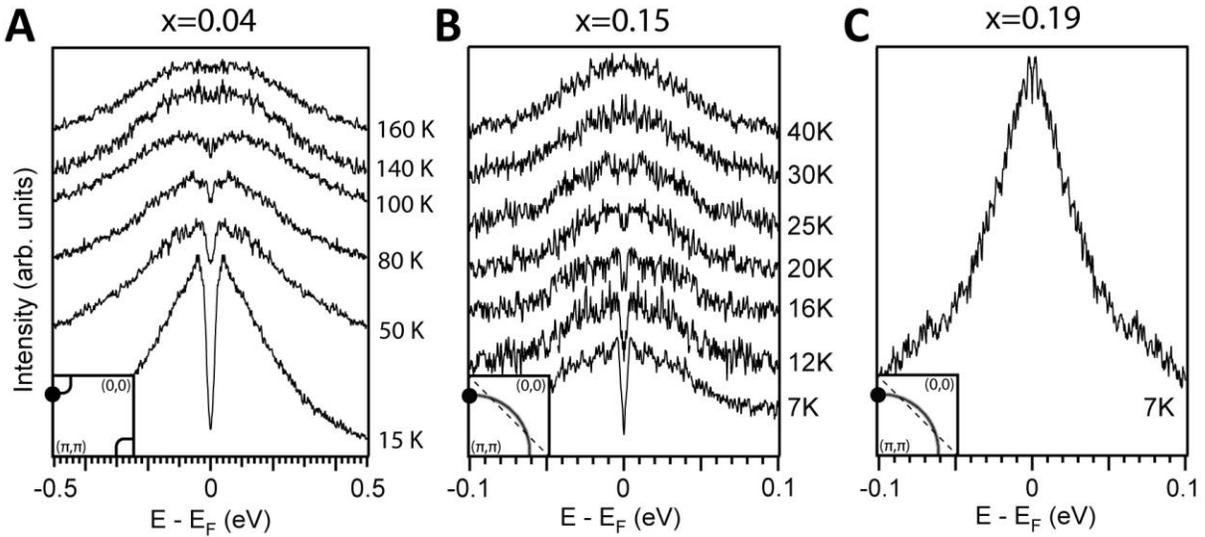

**Fig. S5.**

**Temperature dependence of zone boundary spectra for x = 0.04, x = 0.15, x = 0.19 samples.**
**(A-C)** Symmetrized EDCs measured at the Brillouin zone boundary for x = 0.04 (**A**), x = 0.15 (**B**), and x = 0.19 (**C**). Measurement temperatures for each curve are indicated by the labels. Curves are offset for clarity. The x = 0.19 spectrum is obtained by surface K dosing (see methods), and the doping level is estimated from the Fermi surface volume. For the x = 0.04 sample, the temperature-dependent measurement was performed in the following order: cleaved sample at 50 K, followed by raising temperature to 160 K, then cooled down to 15 K for the lowest temperature measurement.

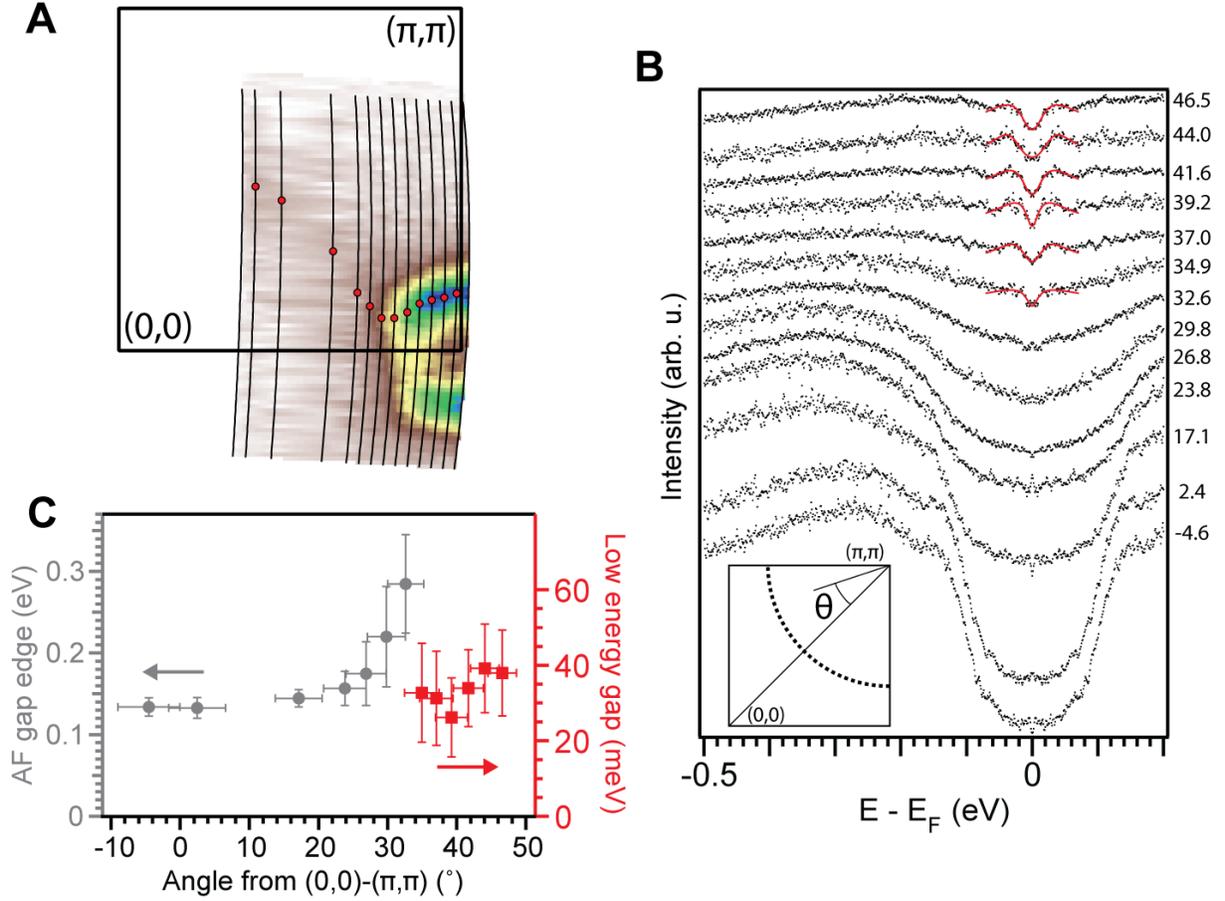

**Fig. S6.**

**Momentum dependence of low energy gap for x = 0.04.** (**A**) Fermi surface mapping of an x=0.04 sample. Black lines indicate the momentum cuts used for mapping. Red dots indicate the Fermi momentum $k_F$. We note that it is difficult to establish the underlying $k_F$ outside of the reconstructed electron pocket due to the very small residual spectral weight. (**B**) EDCs at $k_F$ for the cuts shown in (A). The numbers to the right indicate the angle away from the Brillouin zone diagonal as indicated by $\theta$ in the inset. Blue curves overlaying the low energy spectra for the EDCs near the zone boundary are the fits to the low energy gap. The spectra have a slightly different high energy line shape than that shown in Fig. 3A and supplementary Fig. S6, due to the different cutting directions of the APRES measurement. However, this does not affect the low energy gap measurement significantly. We note that the peak-dip-hump feature from around 50 meV to 200 meV is also observed in the diagonal cutting direction (supplementary Fig. S5C). This feature will be studied in more detail in a future work. (**C**) Angular dependence of the AF gap edge (grey, left) and the low energy gap (red, right) at x = 0.04.

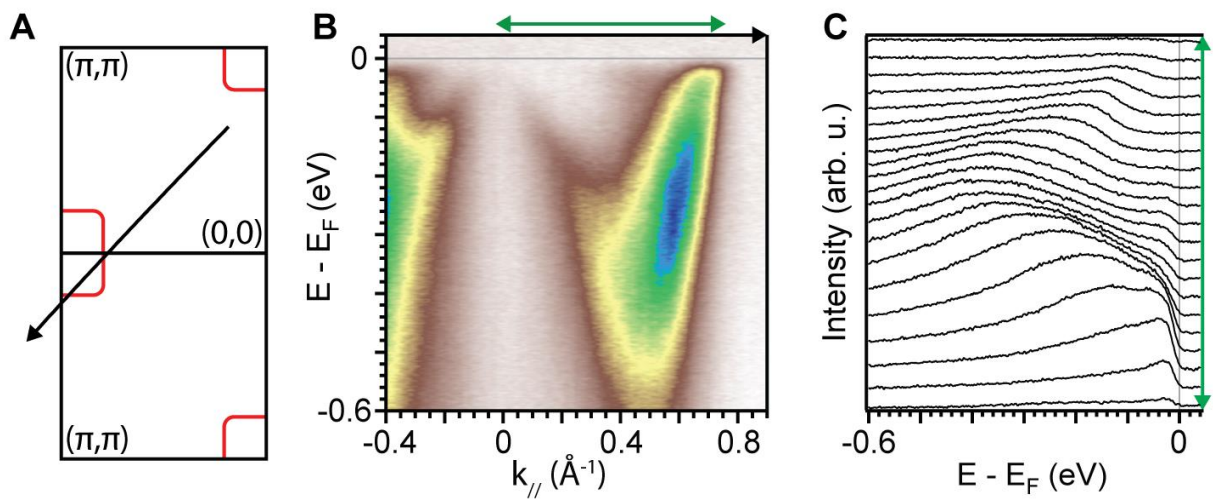

**Fig. S7.**

**Zone boundary spectra for the x = 0.04 sample.** (**A**) Schematic of ½ of the Brillouin zone in NCCO, showing the AF-reconstructed Fermi surfaces. For x = 0.04, the Fermi surface would consist of only the reconstructed electron pockets centered at (0, π) and equivalent momentum points. The black arrow identifies the energy-momentum cut shown in (B). (**B**) Energy-momentum cut across the zone boundary as indicated by the black arrow in (A). Grey line indicates the Fermi level. (**C**) EDCs of the momentum range indicated by the green arrow.

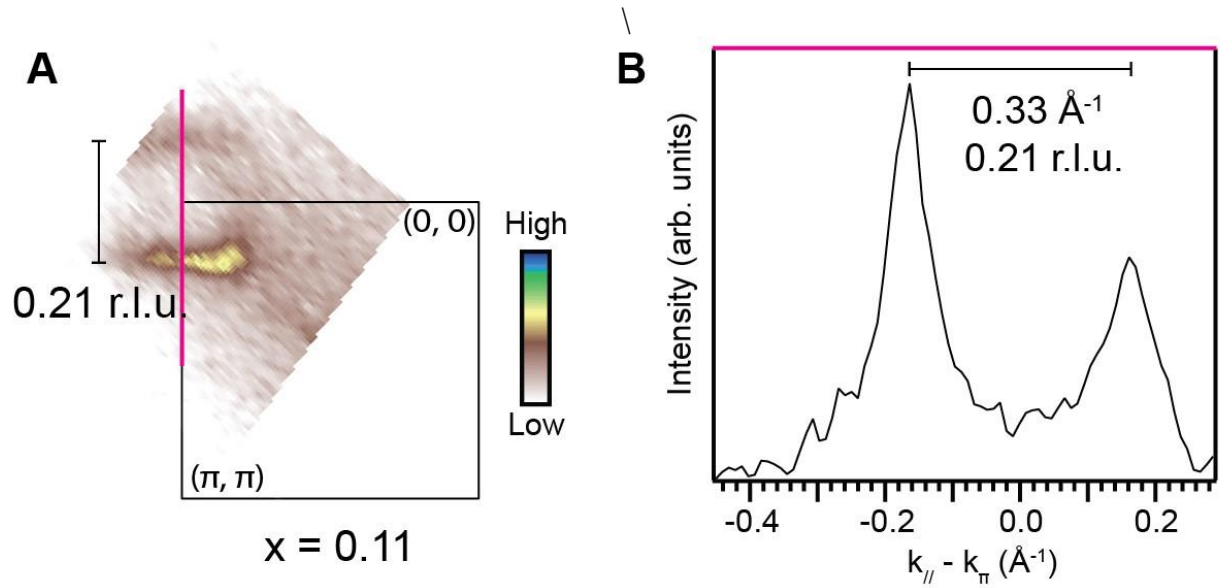

**Fig. S8.**
**Constant energy mapping in the extended zone.** (**A**) Constant energy mapping of an x = 0.11 sample beyond the first Brillouin zone. The color plot is constructed using intensity from ±10 meV of the Fermi energy. (**B**) Momentum distribution curve at EF taken at the cut indicated by the magenta line in (**A**). The separation between the parallel lines of the reconstructed segments is about 0.33 Å$^{-1}$, which is about 0.21 in units of π.

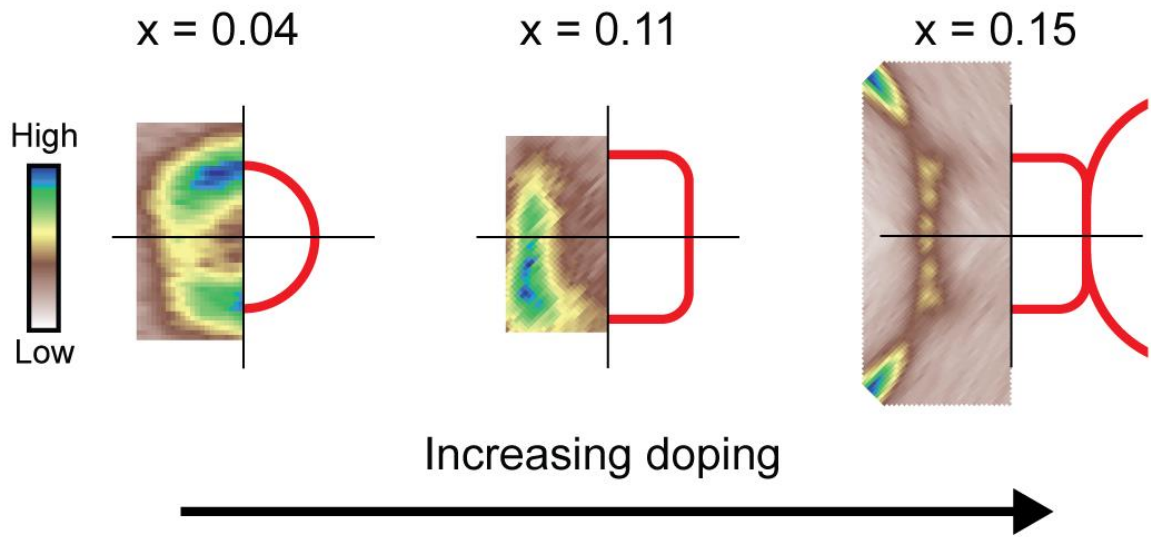

**Fig. S9.**

**Fermi surface shape evolution.** Shape of the Fermi surface near the Brillouin zone boundary as a function of increasing doping from 4% (left) to 11% (middle) to 15% (right). The crossing point of the black lines mark $(0, \pi)$. The red lines indicate the schematic Fermi surface drawings. The data for the 4% and 11% samples are taken in the first and second Brillouin zone. The data for the 15% sample is taken in the first zone and then symmetrized.

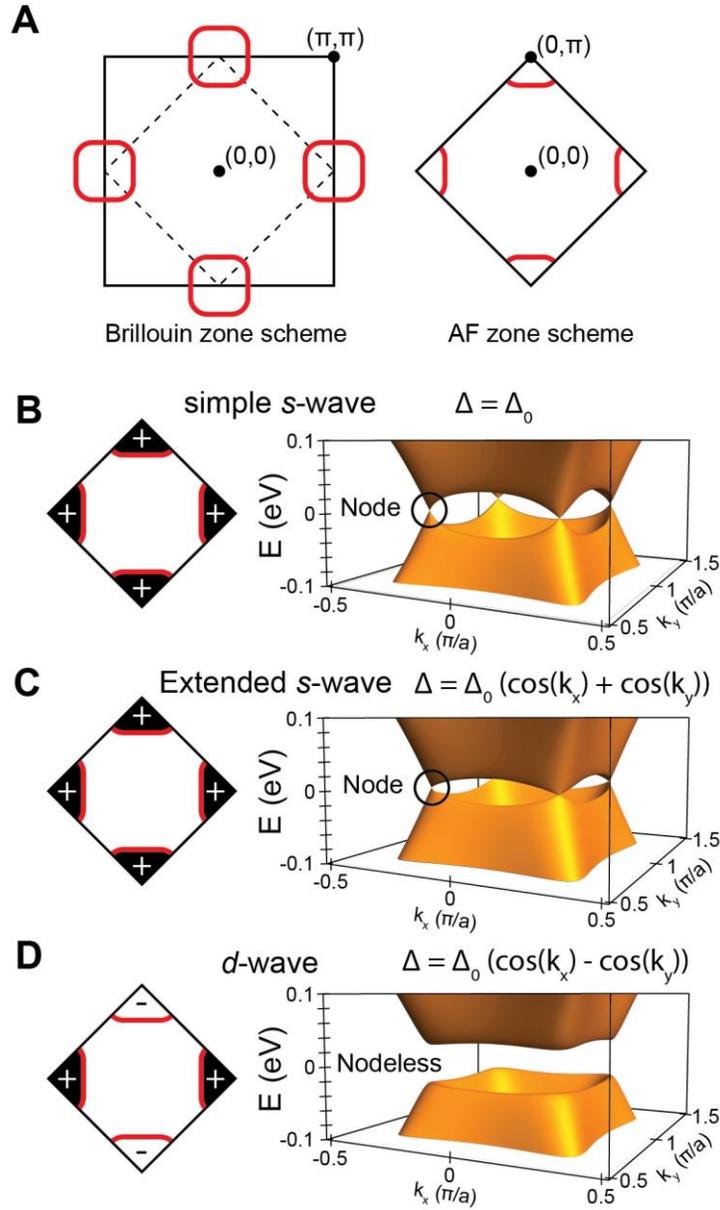

**Fig. S10.**

**Microscopic origins of low energy gap.** (**A**) Schematic of the reconstructed electron pockets in the full Brillouin zone (left) and the AF reconstructed zone (right). In the AF zone, there is only one electron pocket. (**B-D**) Symmetry of the pairing order parameter on the reconstructed electron pocket calculated through real-space nearest-neighbor pairing (see details in the supplementary text), showing a node for the simple s-wave (**B**), extended *s*-wave (**C**), and an approximately isotropic nodeless symmetry for *d*-wave (**D**).

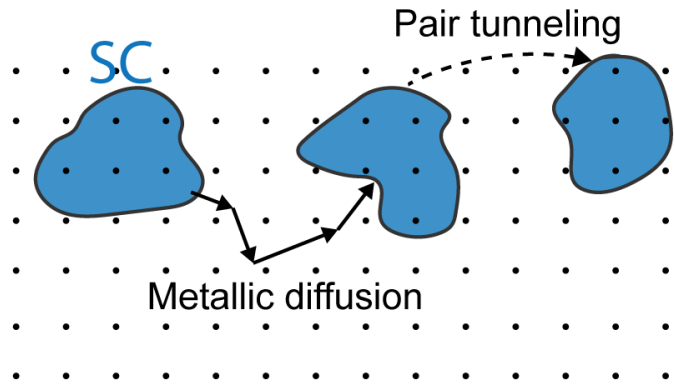

**Fig. S11.**

**Granular superconductivity**. Schematic showing the scenario of fragmented superconducting regions. Transport through the system may be through metallic diffusion or pair tunneling.

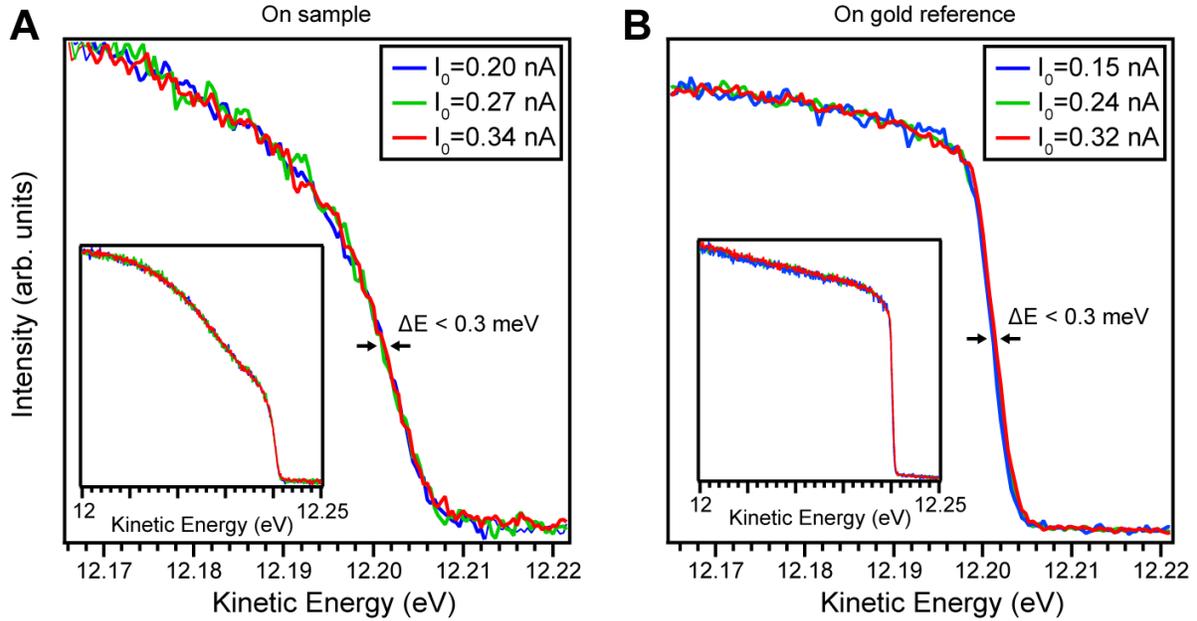

**Fig. S12.**

**Checking for space charging artifacts at SSRL beamline 5-4 by varying the incident beam current $I_0$ measured at a mirror in the X-ray optics.** (**A**) momentum-integrated EDCs near normal emission for a x = 0.15 sample at different $I_0$. The change in the leading edge is noise-limited and less than 0.3 meV. The typical measurement $I_0$ is around 0.2 to 0.25 nA for all measurements at SSRL beamline 5-4. Inset shows the large energy scale spectra used for the EDC normalization process. (**B**) similar $I_0$ dependence on the gold reference, also showing noise-limited leading edge shift less than 0.3 meV. $I_0$ for gold spectra measurements used in $E_F$ extractions is around 0.2 to 0.25 nA. Inset shows the large energy scale spectra used for the EDC normalization process. We note that $I_0$ is not calibrated to an absolute beam current value.

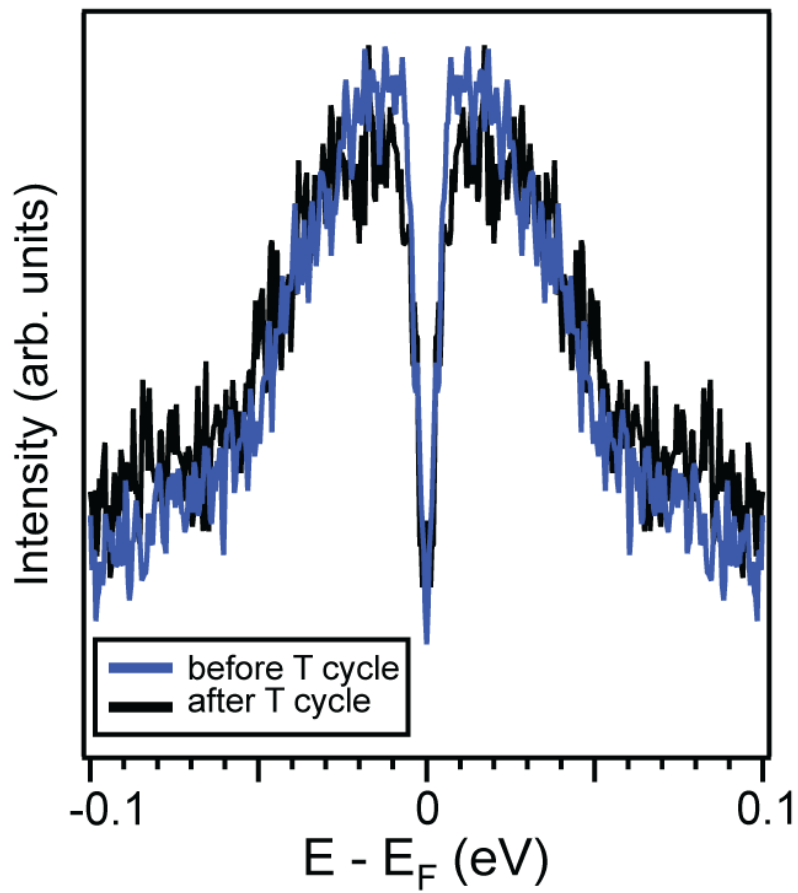

Fig. S13.

**Repeatability of spectra before and after temperature cycling.** Symmetrized EDCs at $k_F$ for the $x = 0.11$ sample, taken before (blue curve) and after (black curve) temperature cycling. The gap and spectra are reproducible, indicating that the gap closing shown in main text Fig. 2C is intrinsic.

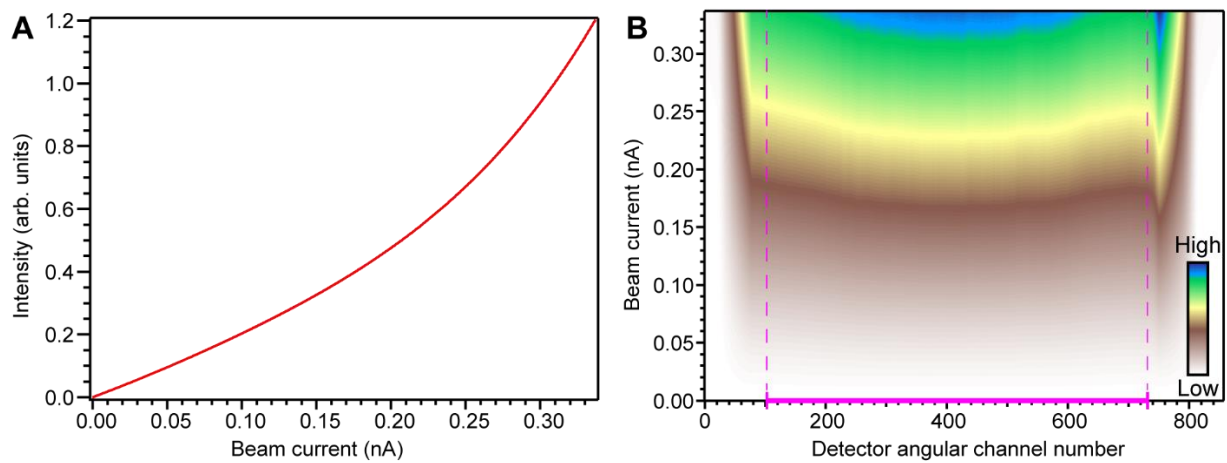

**Fig. S14.**

**Detector nonlinearity correction.** The electron detector can have non-linearities in intensity that can introduce artificial sharpening and downshift of near-$E_F$ features. We correct for this by determining the non-linear scaling using many reference gold spectra at different intensities and applying this scaling to all measured spectra. (**A**) non-linear scaling determined on gold reference, which is fitted with a 5$^{th}$ order polynomial to determine the leading-order non-linear coefficients used for the correction. (**B**) channel by channel non-linear scaling color plot. The outer fringes are detector artifacts. The measurements in this work only use the channels within the magenta bracket, with each channel corrected using the non-linear scaling in (**A**).

## References and not